\RequirePackage{luatex85}
\documentclass[fleqn,10pt]{wlscirep}
\usepackage[utf8]{inputenc}
\usepackage[T1]{fontenc}

\renewcommand{\L}{\text{L}}
\newcommand{\W}{\text{W}}

 \newcommand{\braket}[2]{\langle{#1}|{#2}\rangle}

\usepackage{tikz}                       % Für pgfplots benötigt
\usepackage{pgfplots}                   % Für schöne Plots
\usepgfplotslibrary{fillbetween}
\usetikzlibrary{spy}
\usetikzlibrary{pgfplots.groupplots}
\usetikzlibrary{patterns}

\usetikzlibrary{external}
\tikzset{external/system call={lualatex -shell-escape -file-line-error -halt-on-error 
-interaction=batchmode -jobname "\image" "\texsource"}}
\tikzsetexternalprefix{Bilder/}
\usepackage{shellesc}
\usepackage[miktex]{gnuplottex}

\newcommand{\orcid}[1]{\href{https://orcid.org/#1}{\includegraphics[width=7pt]{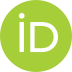}}}

\pgfplotsset{%
  every axis legend/.append style={%
    cells={anchor=west},
    at={(0.96,0.04)},
    anchor=south east,
    font=\scriptsize
  },
  every axis/.append style={%
    yticklabel style={/pgf/number format/fixed zerofill, /pgf/number format/precision=2}
  },
  width= 0.45\textwidth, height=5cm, xmajorgrids=false, xminorgrids=false, minor x tick num=1
}

\pgfplotsset{compat=newest}
 \usepackage{braket}
 \usepackage{upgreek}
 \usepackage{makecell}
\newcommand{\D}{\text{d}}
\def\ba#1\ea{\begin{align}#1\end{align}}																

\title{Quantum free-electron laser oscillator}

\author[1,*]{Peter Kling}
\author[2]{Enno Giese}
\affil[1]{German Aerospace Center (DLR), Institute of Quantum Technologies, Ulm, 89081, Germany}
\affil[2]{Technische Universität Darmstadt, Fachbereich Physik, Institut für Angewandte Physik, Darmstadt, 64289, Germany}
\affil[*]{peter.kling@dlr.de}

\begin{abstract}
If the quantum mechanical recoil of the electron due to its scattering from the undulator and laser fields dominates the dynamics, a regime of the free-electron laser emerges where quantum effects lead to a drastic change in the radiation properties. However, the large interaction length required for a single-pass quantum free-electron laser impedes the experimental realization. The quantum free-electron laser oscillator, proposed in the present article, is a possible scheme to resolve this issue. Here we show that this device features a photon statistics that is closer to a coherent state in comparison to existing classical free-electron lasers. The device can be even operated in such a way that a sub-Poissonian statistics is obtained. Beside the benefit of demonstrating this pure quantum effect, the narrowing of the photon distribution implies reduced intensity fluctuations of the emitted radiation, which in turn lead to decreased noise in imaging experiments or to an enhanced sensitivity in interferometric applications.
\end{abstract}

\begin{document}

\flushbottom
\maketitle
\thispagestyle{empty}

%\noindent Please note: Abbreviations should be introduced at the first mention in the main text – no abbreviations lists. Suggested structure of main text (not enforced) is provided below.

\section*{Introduction}

The quantum free-electron laser~\cite{schroeder,%boni_wigner,
boni06,pio,boni07,bonifacio-basis,boni17,serbeto09,brown17,anisimov18,schaap22,NJP2015,
debus,PRA19,PRR21,PRR23}(Quantum FEL) is a proposed regime of FEL operation where improved features of the emitted radiation are expected. So far, the research focused on single-pass Quantum FELs with the drawback that the required undulator length is rather large and ultimately  reaches a  fundamental limit that is determined by the counteracting effects of multiphoton transitions, spontaneous decay, and space charge~\cite{debus}.

We therefore propose a \textit{Quantum FEL oscillator}, where the emitted radiation of many consecutive electron bunches is stored in a resonator and that can be operated in the low-gain regime with a drastically reduced undulator length compared to high-gain, single-pass FELs. Indeed, the quantum regime requires a high quantum mechanical recoil that translates to an FEL wavelength in the X-ray regime, where up-to-date no suitable high-quality resonators exist. However, in recent years a lot of research was devoted to (classical) X-ray FEL oscillators~\cite{kim08} (XFELO) based on Bragg diffraction from crystals~\cite{colella84} and a large progress was made in this and closely related fields~\cite{huang06,lindberg09,lindberg11,shyvdko11,marcus19,rauer23,margraf23}.     

In the classical limit~\cite{madey1,%madey2,
scully_fel,colson,%kroll,madey3,
sprangle,bnp,huang} many photons are emitted into or absorbed from the laser field ~\cite{mciver} and the discrete nature~\cite{friedman,
%fedorov_rev,becker_jdp,mciver,becker79,becker80,becker80strong,%becker82,
becker83,banacloche,%kurizki,
ciocci} of the electron dynamics is washed out~\cite{carmesin20}. 
However, for a large quantum mechanical recoil $q\equiv 2 \hbar k $ single-photon processes dominate~\cite{NJP2015} and the electron occupies only \emph{two} resonant momentum levels, namely $p\cong q/2$ and $p-q \cong -q/2$. (Note that we consider the Bambini--Renieri frame~\cite{bambi,brs}, where the wavenumbers of the wiggler and laser mode are equal, that is $k_\W=k_\L\equiv k$, and the motion of the electrons is non-relativistic.)
This definition of the Quantum FEL~\cite{NJP2015} can be quantified by the condition
$\alpha_n \equiv g\sqrt{n}/\omega_\text{r}\ll 1$
for the quantum parameter $\alpha_n$, where $g$ denotes the coupling strength of an electron of mass $m$ to the fields, $n$ is the number of laser photons, and $\omega_\text{r}\equiv q^2/(2m\hbar)$ defines the recoil frequency. Moreover, we require that the initial momentum distribution of the electrons, $\rho=\rho(p)$, has to be centered around $p=q/2$ and that its width $\Delta p$ has to be small~\cite{pio,NJP2015}, that is $\Delta p< q$. 

In an oscillator configuration many electron bunches, each consiting of $N$ electrons, are injected with a rate $1/\tau_\text{inj}$ %into the wiggler 
and interact subsequently with the fields inside a cavity which simultaneously stores and damps the generated light field~\cite{scullylamb}. Thus, we identify two contributions of the laser field dynamics, that is (i) Rabi oscillations during the flight time $T$ of electrons through the undulator with the 
momentum-dependent Rabi frequency~\cite{NJP2015,jc}  
\begin{equation}\label{eq:rabi} 
\Omega_n(p) \equiv \sqrt{g^2(n+1)+\omega_\text{r}^2\left(\frac{p}{q}-\frac{1}{2}\right)^2}\ \ ,
\end{equation} 
and (ii) cavity damping which is characterized by the quality $Q$ of the cavity. % and the number $n_\text{th}$ of thermal photons.
With the help of standard methods from laser and micromaser theory~\cite{scullylamb,meschede,lugiato,fili,guzman,schleich} we derive the formal expression
\begin{equation}\label{eq:recurrence} 
P_n=\prod\limits_{n'=1}^n\left[\theta^2
\int\!\!\D p \,\rho(p)\, \text{sinc}^2\left(\Omega_{n'-1}(p)T \right)\right]
\end{equation}
for the photon statistics $P_n$ at steady state, up to a normalization constant. Here we have introduced the pump parameter  $\theta\equiv gT\sqrt{N_a}$  as well as the inverse loss parameter  $N_a\equiv NQ/(\omega_\L \tau_\text{inj})$ with the laser frequency $\omega_\L=ck_\L$ in the laboratory frame.
Compared to its classical counterpart we obtain a narrowed photon distribution in the quantum regime ranging from super- to sub-Posissonian behavior. This narrowing implies reduced intensity fluctuations of the emitted radiation and by that an increase of the signal-to-noise ratio in imaging schemes or to an enhancement in sensitivity in interferometric applications.

%could be exploited in applications where the phase of the emitted radiation is of importance like for example in interferometers [Formulierung/Begründung (n.o.) + falls möglich Anwendungen im Röntgenbereich (Peter)]. 
	  
A shorter undulator length for an oscillator facilitates possible Quantum FEL experiments, for example with respect to the implementation of the necessary optical undulator~\cite{boni07,boni17,debus}, since the focal area and pulse duration of the required high-power laser are limited. However, similar to a single-pass Quantum FEL there are tight bounds on the electron beam quality, especially for the energy spread and the beam emittance. Indeed, the emittance of electrons from (photo-)cathodes cannot fall below an intrinsic thermal limit~\cite{reiser08}. However, advanced schemes like for example ultracold electron sources~\cite{claessens05,franssen19}, that use ionization from ultracold atoms, or plasma wakefield acceleration~\cite{habib23} have the potential to provide the required beam quality in the future.    

%[Thermal limit?, current literature,  ultracold source?]      

\section*{Results}

\begin{figure}[t]
\centering

\begin{minipage}[t]{0.33 \linewidth}
\includegraphics[]{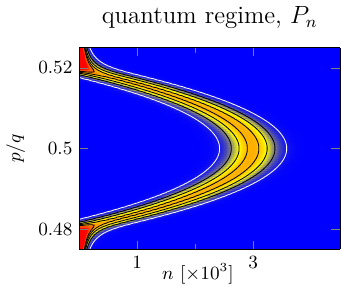}   
\end{minipage}
\begin{minipage}[t]{0.36 \linewidth}
\includegraphics[]{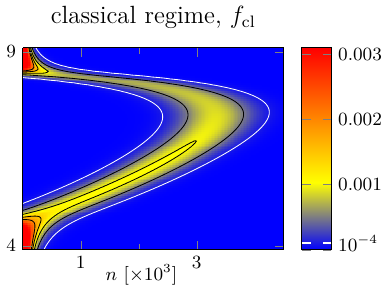}    
\end{minipage}
\begin{minipage}[t]{0.3 \linewidth}    
\includegraphics[]{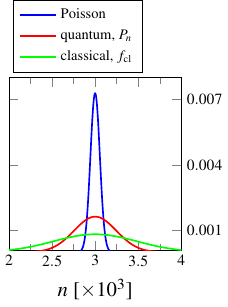}          
\end{minipage}   
\caption{Steady-state photon statistics $P_n$ of an FEL in the quantum regime (left) with $\omega_\text{r}T=20$ and the corresponding distributiuon function $f_\text{cl}$ in the classical regime (middle) with $\omega_\text{r}T=0.2$, both depending on the photon number $n$ and the momentum $p$ of a peaked initial momentum distribution $\rho(p')=\updelta(p'-p)$. In order to compare quantum with classical regime we have chosen in both cases the same values for the relative deviation $\delta=\delta^{\text{(cl)}}=0.05$ of losses from gain and the same mean photon number $ \braket{\hat{n}}=3\cdot 10^3$ (for resonance). %as well as vanishing thermal photons, that is $n_\text{th}=0$ %[TO DO: warum nth null?]. 
This choice of parameters brings us to $N_a=2\cdot 10^4$ in the quantum and $N_a=1.5\cdot 10^3$ in the classical case, respectively, for the inverse loss parameter $N_a\equiv NQ/(\omega_\L \tau_\text{inj})$. We observe that the curve in momentum space of a classical FEL covers multiples of the recoil $q$ and thus is broadened in comparison to its quantum counterpart which is sharply peaked around $p=q/2$. Moreover, the statistics, $P_n$ against $n$, for a Quantum FEL is narrower  as in the classical case which is even more pronounced in the picture on the very right. Here $P_n$ and $f_\text{cl}$ for their respective maximum, at $p=q/2$ and $2kpT/m=\pi$, are compared to a Poisson statistics. Both curves for a small-signal FEL show a super-Poissonian behavior, but due to the additional broadening, Eq.~\eqref{eq:classical}, of $f_\text{cl}$  with $1/(\omega_\text{r}T)\gg 1$ the photon statistics of a  Quantum FEL is closer to the Poisson distribution.}
\label{fig:cl_qu}
\end{figure}

%\subsection*{Quantum vs. classical}

In the following we briefly compare the properties of the Quantum and the classical FEL~\cite{schmueser,becker_pstat,
becker2,banacloche_pstat,%becker_lw,gover_lw,
orszag_lw}. For the time being, we restrict ourselves to the small-signal limit, that is $gT\sqrt{n}\ll 1$ in the quantum regime, leading to a change of the photon number which scales linearly %, respectively quadratic,%
with the initial photon number $n$. %Similar to ordinary laser theory~\cite{scullylamb} we find for resonance, 
A Gaussian approximation~\cite{lugiato} of the photon statistics yields the expression~\cite{diss}
\ba\label{eq:quantum} 
\sigma^2\cong\frac{1}{\delta}\,
\ea
for the Fano factor~\cite{schleich} $\sigma^2\equiv \text{Var}(\hat{n})/\braket{\hat{n}}$, where $\delta$ denotes the relative deviation $\delta \equiv (\mathcal{G}_1-\omega_\L\tau_\text{inj}/Q)/\mathcal{G}_1$
of the losses $\omega_\L\tau_\text{inj}/Q$ from linear gain $\mathcal{G}_1\equiv (gT)^2N $ at $p=q/2$. The FEL is operated above threshold and in the small-signal regime if $0<\delta\ll 1$, resulting in a value for $\sigma^2$ that is larger than unity. Hence, the Quantum FEL shows a super-Poissonian behavior in the small-signal limit.

In the classical regime the small-signal limit is characterized by the Madey gain~\cite{madey1,madey2}
\ba\label{eq:g1_cl} 
\mathcal{G}_1^\text{(cl)}\cong\frac{16}{\pi^3} \, \omega_\text{r}T (gT)^2 N=\frac{16}{\pi^3} \omega_\text{r}T \, \mathcal{G}_1
\ea
for the `classical resonance'  $2kpT/m \cong \pi$ which gives the maximum of a smooth gain curve and thus differs from the sharp peak at $p=q/2$ for a Quantum FEL.
As a consequence, the Fano factor in the classical regime is given by~\cite{becker87,banacloche_pstat,diss}
\ba\label{eq:classical} 
\sigma_\text{cl}^2 \cong \frac{\pi/4}{\omega_\text{r}T}\frac{1}{\delta^{\text{(cl)}}}
\ea 
which can be for example derived~\cite{diss} with an ansatz relying on a Fokker-Planck equation~\cite{risken,haken,orszag_lw}. %and the small-recoil limit.  
Since Eq.~\eqref{eq:classical} implies the classical limit of a small recoil, that is $\omega_\text{r}T\ll 1$, the Fano factor for the classical FEL is much larger than the corresponding value for the Quantum FEL given by Eq.~\eqref{eq:quantum}. In other words, the photon statistics of the Quantum FEL in the small-signal regime is in principle closer to a Poissonian than the statistics of the classical FEL, as illustrated in Figure~\ref{fig:cl_qu}.

%\subsection*{Photon number and variance}
%Topical subheadings are allowed.

\begin{figure}[t]
\centering
   \begin{minipage}[t]{0.45\textwidth}   
\includegraphics[]{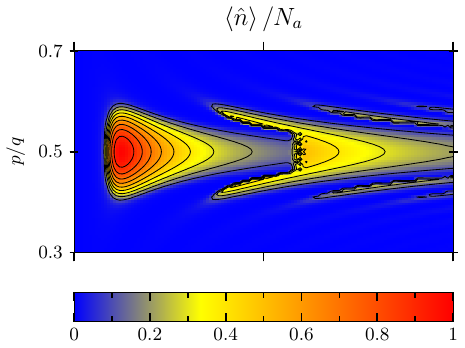}      

\includegraphics[]{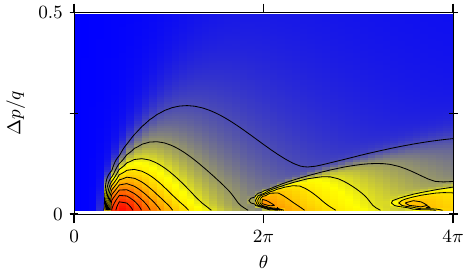}      

\end{minipage}
 \begin{minipage}[t]{0.45 \textwidth}  

\includegraphics[]{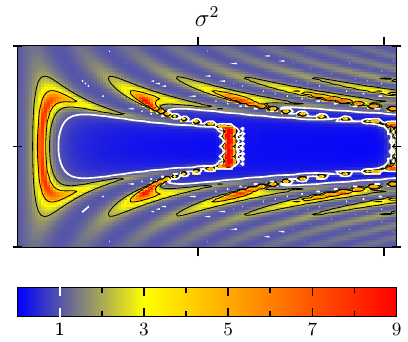}  

\includegraphics[]{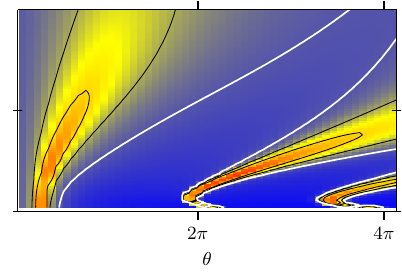} 
  
\end{minipage}  
\caption{Mean photon number $\braket{\hat{n}}$ divided by  the inverse loss parameter $N_a$ (left) and Fano factor $\sigma^2\equiv \text{Var}(\hat{n})/\braket{\hat{n}} $ (right) of a Quantum FEL oscillator for $N_a=150$, %$n_\text{\text{th}}=0$ 
and $\alpha_{N_a}=0.1$. Top: as functions of the pump parameter $\theta$ and the momentum $p$ for an initially peaked momentum distribution  $\rho(p')=\updelta(p'-p)$ of the electrons. Bottom: as functions of $\theta$ and the standard deviation $\Delta p$ of a Gaussian momentum distribution $\rho(p')=(\sqrt{2\pi}\Delta p)^{-1}\exp[{-(p'-q/2)^2/(2\Delta p^2)}]$  centered at $p=q/2$. For momenta $p$ deviating from resonance and increasing values of $\Delta p$, respectively, the mean photon number $\braket{\hat{n}}$ decreases. However, this decrease is more slowly for higher values of the pump parameter $\theta$ since there the interaction between electron and field is stronger.
On the right-hand side, we observe situations with a sub-Poissonian statistics, that is $\sigma^2<1$, which is illustrated by the white contour line at $\sigma^2=1$.
} 
\label{fig:nplussigma}
\end{figure}

In Figure~\ref{fig:nplussigma} 
we present the behavior of the mean value $\braket{\hat{n}}$ and the Fano factor $\sigma^2$ from the photon statistics $P_n$ of a Quantum FEL oscillator given by Eq.~\eqref{eq:recurrence} also beyond the small-signal limit. Only  momenta close to resonance $p=q/2$ or small momentum widths $\Delta p$ lead to a nonzero mean photon number. In atomic scattering this resonance width is known as velocity-selectivity~\cite{giese,giltner,szigeti} and solely caused by a Doppler detuning of the momentum distribution.  
For an efficient operation we even have to go beyond the condition $\Delta p < q$ for the momentum spread $\Delta p$ and  demand for 
\ba\label{eq:vel_sel}
\Delta p < \alpha_n q \ll q \,, 
\ea
where we have estimated the width of the resonance in momentum space by means of Eq.~\eqref{eq:rabi}.
%%%notwendig?
However, increasing the pump parameter $\theta$ leads to a stronger interaction between electrons and fields and at some point we also observe amplification for off-resonant momenta.
 
Another feature analogous to the micromaser~\cite{schleich} is the possibility of a sub-Poissonian photon statistics. %To find a measure for a photon statistics to be super- or sub-Poissonian we define~\cite{schleich}  
%the normalized variance $\sigma^2\equiv\text{Var}(\hat{n})/\braket{\hat{n}}$
%as the ratio of variance and mean value. 
Indeed, the Fano factor takes on values which are smaller than unity, that is $\sigma^2<1$, which is highlighted in Figure~\ref{fig:nplussigma} by the white contour line for $\sigma^2=1$. We observe that this sub-Poisson behavior at resonance $p=q/2$ is not fully destroyed by an increasing momentum spread $\Delta p$. Moreover, we expect that also beyond the small-signal limit the photon distribution in the classical regime is very broad~\cite{banacloche_pstat}.
Hence, we identify the sub-Poissonian photon statistics in the quantum regime as a unique feature of a Quantum FEL.

\subsection*{Experimental challenges}

To this end we discuss the challenges for a possible experimental realization of a Quantum FEL oscillator. For this purpose, we  restrict ourselves again to the small-signal limit, that is $gT\sqrt{n} \ll 1$. A large recoil can be achieved by a small wiggler wavelength  $\lambda_\W$ or a high electron energy $\gamma_0$ which becomes more apparent when 
we write the recoil parameter $\omega_\text{r}T$ in terms of the laboratory frame yielding~\cite{diss}
\ba
1\ll \omega_\text{r}T \propto \gamma_0 \frac{L}{\lambda_\W}\,,
\label{eq:rec_lab}
\ea%=16\pi\gamma_0\frac{L}{1+a_0^2}\frac{\lambda_\text{C}}{\lambda_\text{W}}
where we have introduced the interaction length $L$. % the  undulator parameter $a_0$ and the Compton wavelength $\lambda_\text{C}$ of an electron. 
On the other hand, we demand for a moderately high gain to observe amplification of the
laser field. For a given value of the linear gain $\mathcal{G}_1$ we thus obtain for resonance, $p=q/2$, the relation
\ba 
L=\sqrt{\mathcal{G}_1}\frac{c}{g\sqrt{N}}\propto \frac{\gamma_0^2}{\sqrt{\lambda_\W n_\text{e}}}
\label{eq:length_lab}
\ea %=\sqrt{g_1}\frac{2}{\sqrt{\pi}}\frac{\sqrt{\lambda_\text{C}/r_\text{e}}}{\sqrt{1+a_0^2}}\frac{\gamma_0^2}{\sqrt{\lambda_\text{W}n_\text{e}}}
for the length of the wiggler with $n_\text{e}$ denoting the electron density. 
For increasing values of $\gamma_0$
the wiggler length $L$ grows quadratically while  it only scales with $\lambda_\text{W}^{-1/2}$ with the wiggler wavelength. Hence, for  simultaneously being in the quantum regime and reducing the length of the wiggler we propose to operate a Quantum FEL oscillator with a moderately high electron energy while decreasing the wiggler wavelength. This need for a small wavelength quite naturally forces us to employ an optical undulator~\cite{schlicher,schlicher_ieee,sprangle_lw,steiniger} which has also been proposed previously~\cite{boni17}.     

%In contrast to a SASE FEL, which requires a long undulator covering many gain lengths,
%the undulator length % Eq.~\eqref{eq:length_lab}, 
%necessary for an oscillator is only a fraction of a gain length, due to  $\mathcal{G}_1\ll 1$. However, one faces the problem of constructing a high-quality cavity for X-rays to store the laser field in many passages of electrons.  

\begin{table}
\centering
 \begin{tabular}{llrllr}
 \toprule
 \multicolumn{6}{c}{\textbf{Experimental parameters}}\\
 \cmidrule(r) {1-3}  \cmidrule (l) {4-6}
 laser wavelength  & $\lambda_\L$ (\AA)& $1.0$ & gain bandwidth & 
 $\Gamma$ & $7.45 \cdot 10^{-5} $  \\ 
 undulator wavelength  & $\lambda_\W$ ({\textmu}m)& 1.064 & electron energy & $\gamma_0$ & 51.8\\  
 recoil parameter & $\omega_\text{r}T$ & $2\pi$  & undulator length  & $L$ (mm)& 1.14   \\
 spontaneous emission & $R_\text{sp}L$ & 0.145  & undulator parameter & $a_0$ & 0.0944  \\
space charge & $k_\text{p}L$ & 0.145 &  electron density & $n_\text{e}$ ({\textmu}m$^{-3}$) & $6.39\cdot 10^{4}$    \\ 
linear gain & $\mathcal{G}_1$ & 0.1   & & &  \\
 \cmidrule(r) {1-3}  \cmidrule (l) {4-6}
 \bottomrule
 \end{tabular}
 
\caption[newline]{Proposed fundamental parameters for a Quantum FEL oscillator experiment in the small-signal regime.}
\label{tab:exp}
\end{table}

We have to ensure~\cite{debus} that apart from multiphoton effects the impact of (i) space charge and of (ii) spontaneous emission can be neglected. Hence, we aditionally require that $k_\text{p}L\ll 1 $ as well as $R_\text{sp}L\ll 1$ where we have introduced the plasma wavenumber $k_\text{p}\propto\sqrt{n_\text{e}/\gamma_0^3}$ and the spontaneous decay rate $R_\text{sp}\propto a_0^2/\lambda_\W$ with $a_0$ denoting the undulator parameter. However, these  quantitities are not independent from each other, but are connected   
\begin{equation}
\label{eq:kp_Rs_wr_G}
(k_\text{p}L)^2\cdot (R_\text{sp}L)=\frac{2\alpha_\text{f}}{3}\,
\mathcal{G}_1\cdot (\omega_\text{r}T)
\end{equation}
via the gain $\mathcal{G}_1$ and the fine structure constant $\alpha_\text{f}\approx 1/137$. In contrast to a single-pass FEL~\cite{debus}, we straightforwardy identify a regime, where all conditions can  be simultaneously satisfied. If we for example demand for a gain $\mathcal{G}_1$ of 10\% and set the recoil parameter to $\omega_\text{r}T=2\pi$, Eq.~\eqref{eq:kp_Rs_wr_G} allows for $k_\text{p}L=R_\text{sp}L\approx 0.145$. This more advantageous behavior is due to the low gain which comes hand in hand with a decreased interaction length.%we thus require a high-quality electron beam with a very narrow energy spread, which is challenging for today's experimental capabilities. 
We still have the freedom to adjust two parameters to fully determine the set of fundamental paramaters $\gamma_0$, $n_\text{e}$, $\lambda_\W$, $a_0$, and $\lambda_\L$. From the choice $\lambda_\L=1\,${\AA} and $\lambda_\W=1.064\,$\textmu m we derive the values that are listed on the right-hand side of Tab.~\ref{tab:exp}. % [TO DO: Begründung] 

\begin{figure}[t]
  \centering 
 \begin{minipage}{0.6\textwidth}  
  \includegraphics[]{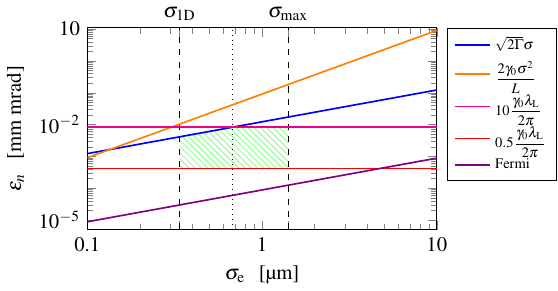}  
 \end{minipage}
 \begin{minipage}{0.35\textwidth}
   \includegraphics[]{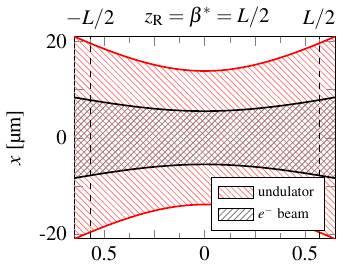} \\
   \includegraphics[]{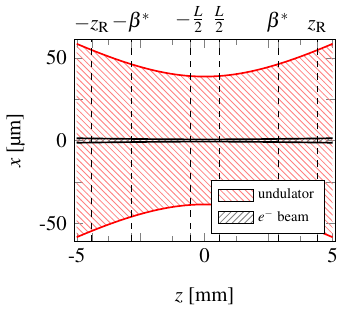} 

 \end{minipage}
  \caption{Experimental challenges for a Quantum FEL oscillator: On the left-hand side present different constraints on the transverse normalized emittance $\epsilon_n$ of the electron beam electron beam as functions of the beam radius $\sigma_\text{e}$. Only parameter pairs within the green shaded area fulfill these conditions leading to a situation with  $\epsilon_n\sim 10^{-2}\,$mm mrad and $\sigma\sim\,${\textmu}m, which is extremely challenging for an experimental realization.     On the right hand side we have shown the longitudinal, $z$, and transversal, $x$, dimensions of the undulator and of the electron beam, when (i) only geometrical considerations are taken into account and when (ii) all conditions are satisfied.    While in the former case, both, the Rayleigh length $z_\text{R}$ of the optical undulator and its counterpart $\beta^*$ for the electron beam coincide wit half of the interaction length $L$, that is $z_\text{R}=\beta^*=L/2$, we observe in the latter case   an increased value of $z_\text{R}$ and the relation $L/2< \beta^* < z_\text{R}$. Hence,  the electron beam occupies only   a small area of the undulator beam.}
  \label{fig:emittance}
\end{figure}

Besides such fundamental constraints we have to take geometrical considerations into account. We assume that the optical undulator  can be described as a Gaussian laser beam with waist $w_0$ and Rayleigh length $z_\text{R}\equiv \pi w_0^2/\lambda_\W $ %(To DO: Beam shape). 
Following other studies~\cite{debus} we require $w_0\geq \sqrt{2\pi} \sigma_\text{e}$, where $\sigma_\text{e}$ denotes the radius of the electron beam. Moreover, the Rayleigh length $z_\text{R}$ has to be at least as long as half of the interaction length, that is $ 2z_\text{R}\geq L$. Similar to the Rayleigh length of a light beam, the quantity $\beta^*\equiv\sigma_\text{e}^2\gamma_0 /\epsilon_n$ describes the typical length for the divergence of an electron beam. The normalized transverse emittance  $\epsilon_n$ is connected to the area of the beam in phase space~\cite{reiser08}. From $2\beta^*\geq L$ we derive~\cite{debus} the condition $\epsilon_n\leq \sigma_\text{e}^2\gamma_0/L$.

We also have to relate the dimensions of the electron beam to the emitted radiation~\cite{debus} characterized by the beam emittance $\lambda_\L/(4\pi)$. A compromise between a strongly diverging laser beam ($\lambda_\L/(4\pi)> \epsilon_n/\gamma_0$) and the loss of transverse coherence ($\lambda_\L/(4\pi)\ll \epsilon_n/\gamma_0$) is typically found in the regime~\cite{debus} 
\begin{equation}
0.5\, \frac{\gamma_0\lambda_\L}{2\pi} < \epsilon_n \leq 10\, \frac{\gamma_0\lambda_\L}{2\pi}\,.
\label{eq:e_trans}
\end{equation}
Indeed, this relation that ensures transverse coherence was originally derived~\cite{saldin2008_coherence,saldin2008_output} for self-amplified spontaneous emission (SASE). However, the radiation in an oscillator also starts up from vacuum and reaches steady state after passages of many electron bunches. There is no mechanism evident inducing transverse coherence and we consequently demand that Eq.~\eqref{eq:e_trans} is also satisfied for an oscillator, assuming transverse coherence for the start-up from vacuum.%PK den Begriff SASE würde ich hier nicht verwenden    

In Tab.~\ref{tab:exp} we have also included the value for the gain bandwidth $\Gamma$. In the small-signal limit the condition from momentum selectivity, Eq.~\eqref{eq:vel_sel}, reads $\Delta p \leq q/(\omega_\text{r}T)$ and we obtain
\begin{equation}
\frac{\Delta \gamma_0}{\gamma_0}\leq\Gamma\equiv \frac{\lambda_\W}{4\pi L}
\end{equation}  
which results in a challenging value for the electron energy spread at the order of $\sim 10^{-4}$. We can relate fluctuations and deviations of several parameters to the gain bandwidth $\Gamma$. For this purpose, we  note the fundamental relation~\cite{debus,ride95}   
\begin{equation}
\label{eq:lambdaL}
\lambda_\L=\frac{\lambda_W}{2\left(1-\cos{\phi}\right)\gamma_0^2}
\left(1+a_0^2+\gamma_0^2\vartheta^2\right)\,,
\end{equation}
where $\vartheta$ is the emission angle and $\phi$ denotes the angle between the propagation directions of  electrons and optical undulator. For example, we find the  conditions $\Delta\lambda_\W/\lambda_\W\leq 2\Gamma$ and $\Delta I_0/I_0\leq 2\Gamma/a_0^2$ for the bandwidth and the intensity flucuations of the undulator laser with intensity  $I_0\propto a_0^2$. Divergence of the electron beam leads to deviations from the head-on geometry ($\phi=\pi$ and $\vartheta=0$) resulting in the condition~\cite{debus} $\varepsilon_n\leq \sigma_\text{e} \sqrt{2\Gamma}$, where we have set $\gamma_0 \epsilon_n = \updelta \phi \sigma_\text{e}$.  

Moreover, one can derive requirements for the available interaction area due to longitudinal and transversal intensity variations of the undulator laser close to the focus, resulting in $\Delta z/ z_\text{R}\leq\sqrt{2\Gamma/a_0^2}$ , and, $\Delta x/w_0 \leq \sqrt{\Gamma/a_0^2}$, respectively~\cite{debus}. For a minimized beam waist we thus derive a maximum radius of $\sigma_\text{max}=[\Gamma/(2a_0^2)]^{1/4}\sqrt{\lambda_\W L}/(2\pi)$ for the electron beam.    

For the sake of completeness, we have to exclude Fermionic effects by demanding~\cite{bonifacio-basis} that the number $N$ of electrons is smaller than the phase space volume of the electron bunch divided by $\hbar^3$ leading to the condition 
\begin{equation}
\epsilon_n > \sigma_\text{e} \sqrt{\left(\frac{\lambda_\text{C}}{2\pi}\right)^3\frac{\pi\gamma_0 n_\text{e}}{\Gamma}}\,,
\end{equation}
where $\lambda_\text{C}$ denotes the Compton wavelength of an electron. We also have to ensure that our one-dimensioanl approach is correct and require~\cite{schmueser} $\sigma_\text{e} > \sigma_{1\text{D}}\equiv \sqrt{L \lambda_\L}$.

We visualize the conditions for the radius $\sigma_\text{e}$ and  the normalized emittance $\epsilon_n$ of the electron beam on the left-hand side of Fig.~\ref{fig:emittance} at hand of the parameters in Tab.~\ref{tab:exp}. The regime which allows for a realization of a Quantum FEL oscillator is indicated by a green shading. We realize that the required beam radius lies in the {\textmu}m range with a normalized emittance at the order of $10^{-2}\,$mm mrad, which are extremely challenging requirements for an experiment. 
On the right-hand side of Fig.~\ref{fig:emittance} we sketch the  longitudinal and transverse dimensions of electron beam and undulator laser, when (i) only the geometrical requirements are met, that is $z_\text{R}=\beta^*=L/2$ (top), and (ii) all conditions are satisfied leading to $z_\text{R}>\beta^*>L/2$ (bottom). We observe that $z_\text{R}$ and $w_0$ increase in the latter case, but at the same time the electron beam occupies only a small area of the undulator beam. To improve the situation, a traveling-wave Thomson-scattering scheme (TWTS) has been proposed~\cite{steiniger,debus}, where the undulator laser has a tilted front and copropagates with the electron beam under an certain angle. While this method relaxes the conditions for the optical undulator, it has no influence on the strict requirements for the electron-beam quality.

\begin{table}
  \centering
    \begin{tabular}{llrllr}
     \toprule
      \multicolumn{3}{c}{\textbf{Electron beam}}&
      \multicolumn{3}{c}{\textbf{Optical undulator}}
      \\
       \cmidrule(r) {1-3}  \cmidrule (l) {4-6} 
        electron energy & $\gamma_0$ & 52 & undulator wavelength & $\lambda_\W$ ({\textmu}m) & 1.064 
         \\
        electron density & $n_\text{e}$ ($\text{{\textmu}m}^{-3}$) & $6.4\cdot 10^{4}$ &
        undulator  parameter & $a_0$ & 0.094 \\
        bunch radius & $\sigma_\text{e}$ ({\textmu}m) & 0.68 & beam waist & $w_0$ ({\textmu}m) &38.6 \\
       trans. norm. emittance & $\epsilon_n$ (mm mrad) & 0.0082 & Rayleigh length & $z_\text{R}$ (mm)
        & 4.4 \\
    beam duration & $\tau_\text{e}$ (ps) & 1.2 & pulse duration & $\tau_0$ (ps) & 7.6 \\ 
    peak current & $I_\text{p}$ (A) & 8.9 & intensity & $I_0$ 
    ($\frac{\text{PW}}{\text{cm}^2}$) & 21.7  \\  
bunch charge & $Q$ (pC) & 11 & laser power & $P_0$ (TW) & 0.5 \\      
energy spread & $\displaystyle\frac{\Delta \gamma_0}{\gamma_0}$ ({\textperthousand}) & 0.075 & undulator  linewidth & $\displaystyle\frac{\Delta\lambda_\W}{\lambda_\W}$ (\textperthousand) &0.15\\     
&&& intensity fluctuations &$\displaystyle\frac{\Delta I_0}{I_0}$ (\%) & 1.7  \\  
       \cmidrule(r) {1-3}  \cmidrule (l) {4-6} 
    \bottomrule
   \end{tabular}
  
 \caption[newline]{Proposed parameters for the electron beam and the optical undulator of a low gain Quantum FEL oscillator based on the values in Tab.~\ref{tab:exp}. The values are chosen such that $10^6$ photons are outcoupled from a cavity with reflectivity $R=95.05$\,\% at the wavelength $\lambda_\L=1$\,{\AA}.% The relative change $\Delta L_\text{cav}/L_\text{cav}$ of the cavity length due to vibrations has to be below 0.15\,{\textperthousand}.
 }
 \label{tab:exp1}
\end{table}  
   
Similar to the Rayleigh length of the undulator, its pulse duration $\tau_0$ has to be larger than twice the interaction length divided by $c$, that is $\tau_0>2L/c\sim 10\,$ps. Further, neglecting slippage implies~\cite{schmueser} that the bunch length $c\tau_\text{e}$ is larger than the number $L/\lambda_W$ of undulator periods times the emitted wavelength $\lambda_\L$ which means for our case that the beam duration $\tau_\text{e}$ has to be larger than  approximately $0.3\,$fs. 
The beam duration is the missing parameter to determine the bunch charge and with that the number of emitted photons. For XFELOs a gain reduction is expected when the beam duration is at the order of the inverse bandwith of the cavity crystals that is typically in the sub-picosecond regime~\cite{lindberg11}. We therefore require a beam duration larger than 1\,ps. A set of possible parameters for electron beam and optical undulator of a Quantum FEL oscillator is listed in Tab.~\ref{tab:exp1}.    
%If we demand for a minimum number~\cite{debus} of $10^5$ outcoupled photons and assume that the cavity mirrors have a reflectivity of 95.05\,\% , we arrive at a minimum beam duration of $\sim 0.1\,$ps.     
  
A detailed study of the x-ray cavity lies outside the scope of the present article. Instead, we briefly summarize in the following the most important relations and conditions: The back and forward reflected FEL pulses in the resonator have to arrive at the same time as the electron bunches at the interaction area. Hence, we require $L_\text{cav}=c/(2f_\text{rep})$ for the cavity length, where we have introduced the repetition rate $f_\text{rep}=1/\tau_\text{inj}$ of the accelerator. In experimental implementations this rate is limited  and consequently the cavity becomes large. For example, a very optimistic repetition rate of 10 MHz implies a cavity length of 15\,m. Despite the small dimensions of the optical undulator, a Quantum FEL oscillator would thus be far away from being a compact device.

 The steady-photon number $n_\text{st}\equiv \braket{\hat{n}}$ can be determined via the relation $n_\text{st}=3\delta\, N/\mathcal{G}$.  
Further, the quality of the resonator is fundamentally tied to its reflectivity $R$ and its length $L_\text{cav}$~\cite{haken} and with $L_\text{cav}=c/(2f_\text{rep})$  we find the relation
\begin{equation}
    \label{eq:reflectivity}
 R=1-\frac{\omega_\L \tau_\text{inj}}{2Q}\,.
\end{equation}
A relative deviation $\delta$ of gain from losses of 1\% (corresponding to $gT\sqrt{n_\text{st}}\approx 0.17$) thus translates to a reflectivity of 95.05\%.
We assume that there are no losses due to absorption so that 4.95 \% of the photons are transmmitted through a thin mirror.  For our set of parameters this ratio translates to $10^6$ outcoupled photons. We note that focusing the cavity mode to the size of the electron beam at its waist $\sim${\textmu}m would lead to a Rayleigh length in the cm-regime which is well above the interaction length. 

To ensure that the cavity is in resonance with the FEL mode we further require that fluctuations in the cavity length $L_\text{cav}$ due to vibrations or analogously effective length changes from timing errors due to fluctuations in the accelerator~\cite{lindberg09} are small. By means of Eq.~\eqref{eq:lambdaL} we derive $\Delta L_\text{cav}/L_\text{cav}\leq 2\Gamma$  which translates to a length change in the mm to cm regime depending on the cavity length. However, the timing error has to be also much smaller than the inverse bandwidth of the cavity which typically leads to a maximum length change in the {\textmu}m regime~\cite{lindberg09}. Moreover, the x rays produce a heat load on the crystals affecting their reflectivity so that a cooling of the mirrors to a temperarture below 100\,K becomes necessary~\cite{kim12}.

\section*{Discussion}

The steady-state photon statistics of a Quantum FEL oscillator is closer to the Poisson distribution of a coherent state and eventually shows a sub-Poissonian behavior. Beside the benefit of demonstrating this pure quantum effect, a narrowed photon statistics would lead to smaller intensity fluctuations.
In imaging application, this reduction implies  for a fixed value of $\langle \hat n \rangle $ a smaller signal-to-noise ratio, which is connected to the inverse Fano factor.
Besides intensity fluctuations limiting imaging, the sensitivity of interferometric applications depends on the photon statistics of the interference pattern and therefore benefits from a narrower photon distribution.

The interaction length of a low-gain FEL oscilator is smaller compared to a single-pass device. As a result, the conditions on multiphoton effects, spontaneous emission, and space charge can be simultanously fulfilled. Although the shorter interaction length also improves the challenges for an optical undulator, the limits due to focusing are still prominent, at least for head-on geometries. However, TWTS schemes~\cite{steiniger} could help to overcome this issue. Unfortunately, a Quantum FEL oscillator can be hardly reduced in size to be a compact device, since the cavity is much longer than the interaction length due to a limited electron bunch repetition rate. For the quality of the electron beam there are very strict criteria (also because the gain
bandwidth for a Quantum FEL oscillator is very small). At least, the shorter interaction length might reduce the effect that the macroscopic beam dynamics~\cite{debus} impairs the beam quality along the undulator.

\section*{Data availablility}
The datasets used and/or analysed during the current study available from the corresponding author on reasonable request.

% \item Else, the parameters for
%the cavity seem similar to that brought forward for a classical XFELO.

%\section*{Methods}

%Topical subheadings are allowed. Authors must ensure that their Methods section includes adequate experimental and characterization data necessary for others in the field to reproduce their work.

%\bibliography{references}

%\noindent LaTeX formats citations and references automatically using the bibliography records in your .bib file, which you can edit via the project menu. Use the cite command for an inline citation, e.g.  \cite{Hao:gidmaps:2014}.

%For data citations of datasets uploaded to e.g. \emph{figshare}, please use the \verb|howpublished| option in the bib entry to specify the platform and the link, as in the \verb|Hao:gidmaps:2014| example in the sample bibliography file.

\section*{Acknowledgements}

We are grateful to W. P. Schleich for his stimulating input and continuing support.
We also thank  C. M. Carmesin, A. Debus, M. A. Efremov,  M. G{\"u}hr,  M. Oppold, R. Sauerbrey, K. Steiniger and A. Wolf  for many fruitful discussions. P.K. acknowledges funding by the German Aerospace Center.

%Acknowledgements should be brief, and should not include thanks to anonymous referees and editors, or effusive comments. Grant or contribution numbers may be acknowledged.

\section*{Author contributions statement}

P.K.: Conceptualization (equal), Data curation (lead), Formal Analysis (equal), Methodology (equal), Validation (supporting), Visualization (lead), Writing – original draft (lead), Writing – review \& editing (supporting);
E.G.: Conceptualization (equal), Data curation (supporting), Formal Analysis (equal), Methodology (equal), Validation (lead), Visualization (supporting), Writing – original draft (supporting), Writing – review \& editing (lead); all authors reviewed the manuscript. 

\section*{Additional information}

The authors declare no competing interests.

%To include, in this order: \textbf{Accession codes} (where applicable); \textbf{Competing interests} (mandatory statement). 

%The corresponding author is responsible for submitting a \href{http://www.nature.com/srep/policies/index.html#competing}{competing interests statement} on behalf of all authors of the paper. This statement must be included in the submitted article file.

%\begin{figure}[ht]
%\centering
%\includegraphics[width=\linewidth]{stream}
%\caption{Legend (350 words max). Example legend text.}
%\label{fig:stream}
%\end{figure}

%\begin{table}[ht]
%\centering
%\begin{tabular}{|l|l|l|}
%\hline
%Condition & n & p \\
%\hline
%A & 5 & 0.1 \\
%\hline
%B & 10 & 0.01 \\
%\hline
%\end{tabular}
%\caption{\label{tab:example}Legend (350 words max). Example legend text.}
%\end{table}

%Figures and tables can be referenced in LaTeX using the ref command.%, e.g. Figure \ref{fig:stream} and Table \ref{tab:example}.

\end{document}